\documentclass[journal]{IEEEtran}

\usepackage[T1]{fontenc}
\usepackage[utf8]{inputenc}
\usepackage{amsmath,amssymb}
\usepackage{cite}

\usepackage{graphicx}
\usepackage{booktabs}
\usepackage{needspace}
\usepackage{array,tabularx,multirow}
\usepackage[caption=false,font=footnotesize]{subfig}
\usepackage{threeparttable}

\usepackage{xcolor}
\usepackage{url}
\usepackage[hidelinks]{hyperref}

\newcommand{\singlescan}{single-scan}
\ExplSyntaxOn
\NewExpandableDocumentCommand{\Singlescan}{}{\text_titlecase_first:n{\singlescan}}
\ExplSyntaxOff

\hypersetup{%
  pdftitle={ARCO: Regression-Anchored Controllable Diffusion for OCT-to-OCTA Translation},%
  pdfauthor={Jinpeng Liao, Tianyu Zhang, and Chunhui Li}%
}
\title{ARCO: Regression-Anchored Controllable Diffusion for OCT-to-OCTA Translation}

\author{Jinpeng Liao, Tianyu Zhang, and Chunhui Li%
  \thanks{J. Liao, T. Zhang, and C. Li are with the Centre for Medical Engineering and Technology, University of Dundee, Dundee DD1 4HN, U.K. J. Liao is now with the UCL Great Ormond Street Institute of Child Health, London, U.K., and T. Zhang is now with the University of Bradford, Bradford, U.K. (Corresponding author: Chunhui Li; e-mail: c.li@dundee.ac.uk.)}}
\begin{document}
\bstctlcite{IEEEcontrol}
\maketitle

\begin{abstract}
Optical coherence tomography (OCT)-angiography (OCTA) requires repeated B-scans at the same location to extract the microvasculature from the signal changes between them, which lengthens acquisition and increases sensitivity to motion.
Deep-learning OCT-to-OCTA translation overcomes this constraint using only \singlescan{} structural OCT. However, regression models blur uncertain vessels, and diffusion models add vascular detail but can hallucinate vessels absent from the acquired reference.
Refining a frozen regression anchor with a joint five-B-scan diffusion bridge, ARCO gates each correction by the anchor's error predicted by an evidence model, so one control value sets the trade-off between vessel recovery and hallucination, without retraining.
On 50 test volumes (nine with disease) from five oral sites, ARCO reached 84.9\% B-scan completeness at 6.4\% hallucination, compared with the regression baseline VET (61.7\% at 6.8\%).
In en-face analyses, ARCO had lower hallucination (23.0\% vs.\ 25.5\%--51.6\%) and higher completeness (75.3\% vs.\ 49.6\%--71.4\%) than all seven baselines.
In zero-shot inference on 10 skin-hand volumes, ARCO had lower B-scan hallucination than every baseline (23.3\% vs.\ 38.3\%--54.8\%), indicating that its hallucination control transfers to skin.
On the retinal OCTA-500 dataset, the retrained ARCO reached a full-projection structural similarity index (SSIM) of 0.626 and 0.582 on the 3M and 6M subsets, compared with 0.610 and 0.570 for the XOCT model.
By reconstructing OCTA from fast \singlescan{} OCT, ARCO could reduce patient burden and make controllable vascular contrast practical in oral medicine, dermatology and ophthalmology, where repeated scanning is slow.
\end{abstract}

\section{Introduction}\label{sec:introduction}

\begin{figure}[!t]
  \centering
  \includegraphics[width=\columnwidth]{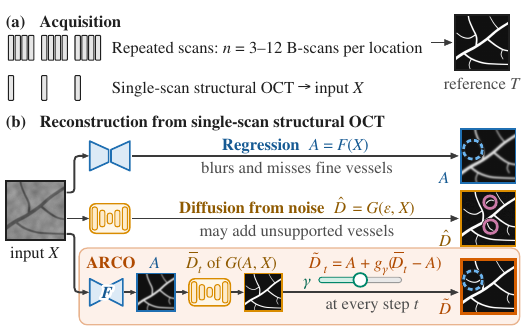}%
  \caption{%
    \Singlescan{} OCT-to-OCTA translation. All panels are schematic.
    (a) Repeated scans give the reference \(T\), and a single scan gives the input \(X\).
    (b) Regression misses fine vessels (dashed circle), and a diffusion sampler \(G\) started from noise adds unsupported ones (solid circles).
    ARCO starts \(G\) from the anchor \(A\) and gates every step with \(\gamma\), shown at \(\gamma=1\).%
  }
  \label{fig:overview}
\end{figure}

Optical coherence tomography (OCT)-angiography (OCTA) extracts microvascular signals from repeated OCT B-scans at the same location (Fig.~\ref{fig:overview}a).
The repeats lengthen the scan and make it sensitive to motion, a burden shared by ophthalmic~\cite{spaide2015artifacts}, skin~\cite{deegan2019skin} and oral OCTA and heaviest intraorally, where a hand-held probe must reach sites such as the floor of the mouth and the tongue~\cite{liao2023ufct,liao2025imast}.
OCT-to-OCTA translation removes the repeats by predicting the microangiogram from \singlescan{} structural OCT (one B-scan per slow-axis location)~\cite{lee2019flow,liao2024vet}.

Among these deep-learning translators~\cite{lee2019flow,liao2024vet,transpro2024,xoct2025}, regression models average over uncertain fine structures, resulting in the loss of small vessels.
Diffusion models instead generate sharp, vessel-rich images~\cite{zhou2024ddbm,arslan2025selfrdb,flowoct2026}, but are prone to hallucinating vessels that are absent from the acquired reference~\cite{kim2024local}. Generative models can refine a regression output~\cite{whang2022stochastic,ohayon2025pmrf}, but the refinement is applied globally and is not spatially adaptive. These limitations motivate a diffusion model that preserves rich local vascular detail and suppresses hallucination.

We propose ARCO, a regression-anchored controllable diffusion model that recovers OCTA from single-scan structural OCT, keeping the regression output as an explicit anchor and refining it with a diffusion bridge (Fig.~\ref{fig:overview}b).
During sampling, an evidence model estimates the pixel-wise error of the anchor, which defines a gate \(g_\gamma\in[0,1]\) that weights the correction.
A single control value \(\gamma\) scales this gate and thereby sets the trade-off between vessel recovery and hallucination without retraining, with the anchor recovered at \(\gamma=0\) and the ungated bridge as \(\gamma\to\infty\).
An ablation study of the gate confirms its effectiveness in suppressing hallucination.
With this design, ARCO reconstructs the fine vessel details that regression loses while keeping hallucination under explicit control.

Our contributions are as follows.
(1) We formulate single-scan OCTA reconstruction as the controlled correction of a regression anchor.
(2) We adapt an hourglass diffusion transformer as the bridge denoiser, which predicts five neighboring B-scans jointly to exploit vessel continuity across them.
(3) We propose a reference-based evaluation that scores hallucination and completeness separately, without vessel annotation.
(4) We validate ARCO on oral, retinal and skin (zero-shot inference) data in comparison with up to seven baselines.%
\section{Related Work}\label{sec:related-work}

\subsection{OCT-to-OCTA Translation}

Regression translators learn the mapping with a pixel-wise loss.
Lee \emph{et al.} trained a U-shaped network with a mean-squared-error (MSE) loss on retinal data~\cite{lee2019flow}, and the vasculature extraction transformer (VET) applied the same objective to skin and oral data~\cite{liao2024vet}.
Regression needs no annotation, but it can miss the finest vessels, and Lee \emph{et al.} reported fewer third- and fourth-order vessels than in the reference OCTA~\cite{lee2019flow}.

To recover this detail, TransPro and XOCT add adversarial training together with a vessel-segmentation loss~\cite{transpro2024} or layer-wise supervision~\cite{xoct2025}.
Both require retinal annotations, and an independent evaluation found that both lose the fine capillary network~\cite{chertok2026defense}.
Diffusion and flow-matching translators face a trade-off between detail and hallucination.
A latent Brownian bridge had lower clinician-rated hallucination than an adversarial model but smoothed fine vascular edges~\cite{badhon2025octa}, and FlowOCT applies flow matching to wavelet coefficients and was evaluated on OCTA-500 alone~\cite{flowoct2026}.
Apart from VET, these translators were validated on retinal data, and, to our knowledge, none has reported high vessel completeness together with low hallucination.%
\subsection{Anchored Refinement and Sampling-Time Control}

A vanilla diffusion model generates the image from noise and is prone to structural hallucination~\cite{kim2024local}.
Diffusion bridges instead start sampling from the source image, which serves as an informative prior~\cite{zhou2024ddbm,dbim2025}.
Implicit sampling accelerates them~\cite{dbim2025}, and SelfRDB extends them to medical image translation~\cite{arslan2025selfrdb}.
A deterministic estimate of the target can be a closer starting point than the source image~\cite{ohayon2025pmrf}.
Predict-and-refine models build on such an estimate, with a diffusion model that learns its residual~\cite{whang2022stochastic} or a rectified flow that transports it to the image distribution~\cite{ohayon2025pmrf}, and MADM starts sampling from a CNN prior image for PET~\cite{chen2025madm}.
Their sampling settings trade distortion for perceptual quality~\cite{whang2022stochastic,ohayon2025pmrf}, consistent with the distortion--perception trade-off~\cite{freirich2021dp}.
However, these settings are global, so the correction cannot be confined to the regions of large estimation error. Spatially adaptive sampling targets this limitation.
In natural-image super-resolution, an uncertainty map of a regression output guides a latent diffusion model to restore detail in uncertain regions~\cite{wang2026ugdiff}, and Local Diffusion~\cite{kim2024local} generates separately inside and outside detected out-of-distribution regions to reduce structural hallucination.
Both start from noise and have not been evaluated on microangiographic images, and the methods that build on a deterministic estimate apply the correction globally.%
\subsection{Error Estimation and Hallucination Metrics}

A spatially adaptive correction requires a per-pixel estimate of the regression error.
A regression network can predict a per-pixel variance together with its output, learned from the regression loss alone~\cite{kendall2017uncertainty}.
In diffusion models, the spread of repeated samples is often interpreted as uncertainty, but it may reflect replicated acquisition noise instead of epistemic uncertainty~\cite{rassmann2026yoda}.
Measuring hallucination requires a reference that shows which structures are real.
The Hallucination Index measures the distance between the distribution of reconstructions and a hallucination-free reference distribution~\cite{tivnan2024hi}, but it needs multiple stochastic reconstructions of each case and was demonstrated on simulated data.
For vessel-like tubular structures, clDice merges the false and the missed parts into one harmonic mean~\cite{shit2021cldice}.
Neither metric reports hallucination and completeness separately, as a comparison of translators along this trade-off requires.%
\section{Method}\label{sec:method}

\begin{figure*}[!t]
  \centering
  \includegraphics[width=0.90\textwidth]{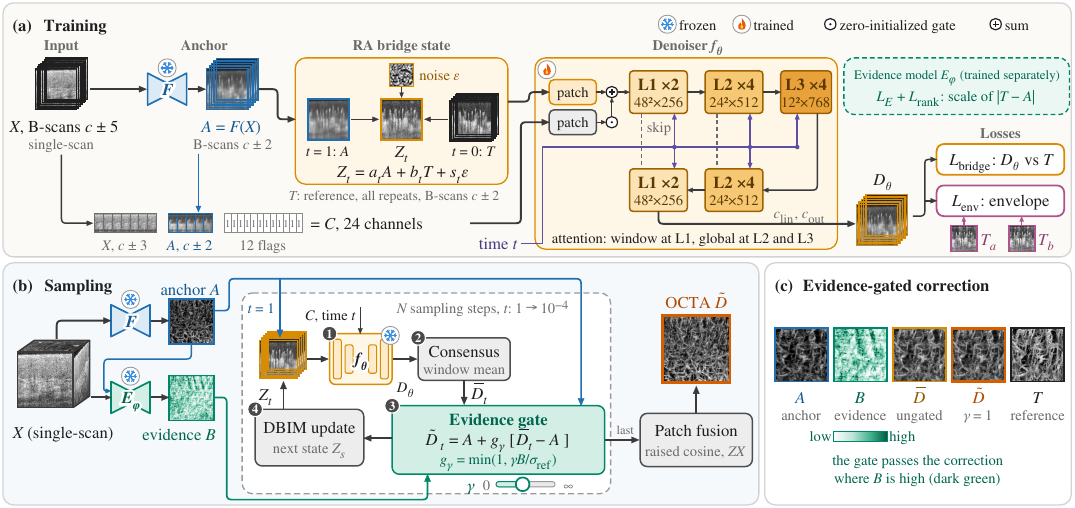}%
  \caption{%
    ARCO training, sampling and evidence-gated correction.
    (a) Training. The frozen regression network \(F\) gives the anchor \(A\), and the diffusion denoiser \(f_\theta\) predicts the clean estimate \(D_\theta\) from the bridge state \(Z_t\), the condition \(C\) and the time \(t\), under the bridge loss and the envelope loss with the half references \(T_a,T_b\).
    (b) Sampling starts from \(A\), and at each sampling step the gate \(g_\gamma\) weights the correction of \(A\) according to the evidence map \(B\) and the control value \(\gamma\).
    (c) Effect of the gate on en-face patches (maximum intensity projection).
    Images show one validation volume of the labial mucosa.%
  }
  \label{fig:framework}
\end{figure*}

ARCO comprises three networks (Fig.~\ref{fig:framework}).
A regression network \(F\) gives the anchor as \(A=F(X)\), a five-B-scan diffusion denoiser trained on a regression-anchored (RA) bridge predicts its correction, and a separately trained, frozen evidence model \(E_\phi\) gates this correction during sampling.
ARCO extends the bridge~\cite{zhou2024ddbm} and sampler~\cite{dbim2025} by gating the correction at every sampling step with the predicted error of the anchor.

\subsection{Inputs and Regression Anchor}\label{sec:anchor}

The RA bridge starts from a regression anchor (Fig.~\ref{fig:framework}a).
Let \(X\) denote the \singlescan{} structural OCT obtained from the first repeat and \(T\) the acquired OCTA reference formed from all repeats.
The frozen regression network \(F\) predicts each B-scan of the anchor $A=F(X)$ from seven neighboring B-scans of \(X\).
For the five B-scans \(c\pm2\) around a center \(c\), the denoiser receives the bridge state \(Z_t\) and a condition \(C\), and it predicts the clean estimate \(D_\theta\) of the same five B-scans.
The condition \(C\) stacks the seven B-scans \(c\pm3\) of \(X\), the five B-scans \(c\pm2\) of \(A\) and their twelve validity flags (1 inside the volume, 0 outside) into 24 channels.%
\subsection{Regression-Anchored Bridge}\label{sec:bridge}

Following the variance-exploding (VE) bridge~\cite{zhou2024ddbm}, the RA bridge connects the reference \(T\) at $t=0$ to the anchor \(A\) at $t=1$, so that sampling starts from \(A\) and the denoiser learns the residual between \(A\) and \(T\).
Its state is
  \begin{align}
    Z_t &= a_t A+b_t T+s_t\epsilon, \label{eq:bridge-state}\\
    a_t &= t^2,\quad b_t=1-t^2,\quad s_t^2=a_tb_t, \label{eq:bridge-schedule}
  \end{align}
where \(\epsilon\) is standard Gaussian noise. Its preconditioning~\cite{zhou2024ddbm,dbim2025} expresses the clean estimate through the network \(f_\theta\) as
  \begin{equation}
    \label{eq:bridge-clean}
    \begin{split}
      D_\theta &= \mu_T+c_{\mathrm{lin}}(Z_t-m_t)\\
      &\quad+c_{\mathrm{out}}f_\theta\!\left(
      c_{\mathrm{in}}(Z_t-m_t),C,250\log t\right).
    \end{split}
  \end{equation}
Let \((\mu_T,v_T)\) and \((\mu_A,v_A)\) be the mean and variance of \(T\) and \(A\) on the paired training data, and \(v_{TA}\) their covariance.
With \(\Delta=v_Tv_A-v_{TA}^2>0\), the coefficients are
  \begin{align}
    m_t &= a_t\mu_A+b_t\mu_T, \label{eq:bridge-mean}\\
    V_t &= a_t^2v_A+b_t^2v_T+2a_tb_tv_{TA}+s_t^2, \label{eq:bridge-variance}\\
    c_{\mathrm{in}} &= V_t^{-1/2},\qquad
    c_{\mathrm{lin}}=\frac{b_tv_T+a_tv_{TA}}{V_t}, \label{eq:bridge-input-skip}\\
    c_{\mathrm{out}}^2 &= \frac{a_t^2\Delta+v_Ts_t^2}{V_t}. \label{eq:bridge-output-scale}
  \end{align}
These coefficients give the input of \(f_\theta\) and its training target unit variance at every \(t\).
The time input \(250\log t\) follows the implementation of~\cite{zhou2024ddbm}.
Since sampling starts from \(Z_1=A\), training sets \(t=1\) with probability 0.1 and otherwise samples \(t\sim\mathcal U[10^{-4},1-10^{-4}]\).
The bridge loss weights the error of the clean estimate over time and noise,
  \begin{equation}
    \label{eq:loss-bridge}
    L_{\mathrm{bridge}}=
    \mathbb{E}_{t,\epsilon}\big[\tfrac{\lambda_t}{|\Omega|}\sum\nolimits_{x\in\Omega}\omega_x\,(D_\theta-T)_x^2\big],
  \end{equation}
where \(\lambda_t=c_{\mathrm{out}}^{-2}\) equalizes the effective loss weight over \(t\), as in~\cite{zhou2024ddbm}, \(\Omega\) is the set of target pixels, and the weight \(\omega_x\) is zero on invalid pixels and otherwise the inverse of the number of five-B-scan windows that contain the B-scan of \(x\).

The repeats of the scan further provide a regularizer that penalizes flow predicted in the background.
With a scan repeat count of at least four, two disjoint groups of repeats form the half references \(T_a,T_b\), and flow predicted where both halves are background is treated as hallucination.
The envelope loss penalizes such flow above the upper envelope \(T_{ab}=\max(T_a,T_b)\),
  \begin{align}
    w &= h\,\mathbf{1}[T_{ab}<\vartheta_T], \label{eq:envelope-weight}\\
    L_{\mathrm{env}} &=\mathbb{E}_{t,\epsilon}\big[\tfrac{\lambda_t}{|\Omega|}\sum\nolimits_{x\in\Omega}\omega_x w_x\,[D_\theta-T_{ab}-\delta]_+^2\big], \label{eq:loss-envelope}\\
    L_{\mathrm{total}} &=L_{\mathrm{bridge}}+L_{\mathrm{env}}, \label{eq:loss-total}
  \end{align}
where \(h\) indicates that the data has half references, \(\vartheta_T\) is the per-B-scan Otsu threshold of \(T\) over its valid pixels, \([u]_+=\max(0,u)\) retains the excess of the prediction over the envelope, and the margin \(\delta=1.69575\) (normalized target units) is the 90th percentile of \(|T_a-T_b|\) over the background pixels of the training volumes.%
\subsection{Five-B-Scan Denoiser}\label{sec:joint-sampling}

To exploit vessel continuity across B-scans, the denoiser \(f_\theta\) receives the bridge state \(Z_t\), the condition \(C\) and the time \(t\), and predicts five B-scans jointly (Fig.~\ref{fig:framework}a).
It adapts the hourglass diffusion transformer (HDiT)~\cite{crowson2024hdit}, with \(4\times4\) patches that divide a \(192\times192\) input into \(48\times48\) tokens, a three-level encoder (widths L1: 256, L2: 512 and L3: 768), a symmetric decoder with skip connections, window attention at the first level and global attention below.
The tokens of \(C\) are produced by a separate patch embedding and added to those of \(Z_t\) through a zero-initialized gate, and the embedding of \(t\) modulates every block.%
\subsection{Evidence-Gated Sampling}\label{sec:evidence-control}

Sampling refines the anchor according to its predicted error (Fig.~\ref{fig:framework}b).
The error is the residual \(T-A\) between the anchor and the reference \(T\) that repeated scans would give, and the evidence model \(E_\phi\) predicts its scale \(\hat b\).
The inputs of \(E_\phi\) are variables available in one \singlescan{} volume (the condition \(C\), the depth below the tissue surface, the local variance of the anchor and the acquisition protocol).
The protocol is used since its settings (the repeat count, the eigen-decomposition rank discarded in forming \(T\) and the sample probe) affect the vascular signal in \(T\) and thereby the residual.
\(E_\phi\) is trained as a heteroscedastic regressor~\cite{kendall2017uncertainty} by minimizing, with equal weight, the Laplace negative log-likelihood of the residual with scale \(\hat b\) and a ranking loss that encourages \(\hat b\) to follow the ranking of \(|T-A|\),
  \begin{align}
    L_E&=\tfrac{1}{|\Omega|}\sum\nolimits_{x\in\Omega}\omega_x\big(|T-A|/\hat b+\log\hat b\big)_x, \label{eq:loss-evidence}\\
    L_{\mathrm{rank}}&=\tfrac{1}{|\Omega|}\sum\nolimits_{k\in\mathcal S}\omega_k\min\!\big(1,|d_k|/s_{\mathrm{train}}\big)\nonumber\\
    &\qquad\times\log\!\big(1+e^{-\operatorname{sgn}(d_k)\ell_k}\big), \label{eq:loss-evidence-rank}
  \end{align}
where \(d_k=|T_k-A_k|-|T_{\pi(k)}-A_{\pi(k)}|\) and \(\ell_k=\log\hat b_k-\log\hat b_{\pi(k)}\) compare pixel \(k\) with a partner \(\pi(k)\), the pixel set \(\mathcal S\) excludes pairs with \(d_k=0\), and \(s_{\mathrm{train}}\approx0.55193\) (normalized target units) is the mean of \(|T-A|\) over the training volumes.
The partner is \(k\) shifted cyclically by half the valid pixels of its B-scan in the 192 pixels below the tissue surface.
Because \(E_\phi\) predicts the scale of the residual (\(T-A\)), its output indicates where the anchor needs correction and serves as the evidence for the gate.
To prevent hallucination in regions that the anchor already reconstructs accurately, the gate suppresses the diffusion correction where the predicted error is small.

At each of the \(N\) sampling steps, the denoiser predicts a clean estimate for every window of five consecutive B-scans of the state \(Z_t\) (marked 1 in Fig.~\ref{fig:framework}b).
The windows advance along the slow axis by a stride of one to five B-scans.
The consensus \(\overline D_t\) of each B-scan is the mean of its estimates from the windows that contain it (marked 2), up to five windows at a stride of one and a single window at a stride of five.
The evidence map \(B(x)\) is the predicted scale \(\hat b\), averaged for each B-scan over its windows, and is computed once per volume.
The evidence gate (marked 3) weights the correction of the anchor by this map and gives the gated estimate
  \begin{align}
    g_\gamma(x)&=\min\!\left(1,\gamma B(x)/\sigma_{\mathrm{ref}}\right), \label{eq:control-gate}\\
    \widetilde D_t(x)&=A(x)+g_\gamma(x)[\overline D_t(x)-A(x)], \label{eq:control-clean}
  \end{align}
where \(\sigma_{\mathrm{ref}}\) is the median of \(B\) over the valid pixels of the training volumes, and the control value \(\gamma\geq0\) sets how far the gated estimate may depart from the anchor.
Where \(B\) is large, the gate passes the full correction \(\overline D_t-A\), and where \(B\) is small, the gated estimate stays close to \(A\) (Fig.~\ref{fig:framework}c).
At \(\gamma=1\), the full correction applies wherever \(B\) reaches \(\sigma_{\mathrm{ref}}\).
At \(\gamma=0\), every gated estimate equals \(A\) and sampling returns the anchor, and as \(\gamma\to\infty\), the gate equals one everywhere, so sampling runs without the evidence gate (the ungated bridge).

The gated estimate then enters the deterministic update of the diffusion bridge implicit model (DBIM)~\cite{dbim2025} (marked 4), whose rule from time \(t\) to the next time \(s<t\) is
  \begin{align}
    \hat\epsilon_t&=(Z_t-a_tA-b_t\widetilde D_t)/s_t, \label{eq:sampling-noise}\\
    Z_s&=a_sA+b_s\widetilde D_t+s_s\hat\epsilon_t, \label{eq:sampling-update}
  \end{align}
where \(\hat\epsilon_t\) is the predicted noise of the state \(Z_t\).
At \(t=1\), the state is the anchor without noise (\(s_1=0\)) and Eq.~(\ref{eq:sampling-noise}) does not apply, so the first gated estimate \(\widetilde D_1\) initializes the bridge state of Eq.~(\ref{eq:bridge-state}) as \(Z_u=a_uA+b_u\widetilde D_1+s_u\xi\) at \(u=1-10^{-4}\), with standard Gaussian noise \(\xi\).
With \(N\geq3\), the other \(N-1\) sampling steps follow uniformly spaced times from \(u\) to \(10^{-4}\), the limits of the training range of \(t\), and the last gated estimate is the output.
Because \(\widetilde D_t\) interpolates between \(A\) and \(\overline D_t\) at every pixel, each state of Eq.~(\ref{eq:sampling-update}) remains a bridge state of Eq.~(\ref{eq:bridge-state}) with \(\widetilde D_t\) in place of \(T\).%
\section{Experiments}\label{sec:experiments}

\subsection{Data}\label{sec:cohorts}

ARCO is developed and tested on three datasets (Table~\ref{tab:data}), an oral dataset for training and testing, the public retinal OCTA-500 dataset~\cite{li2024octa500} for evaluation on the retina and comparison with published methods, and a skin-hand dataset for zero-shot inference at another anatomical site.
The oral dataset comprises 162 training, 34 validation and 50 (41 healthy and nine disease) test volumes.
Volunteers were recruited in four acquisition projects run between May 2022 and January 2024, and the disease volumes show oral lesions, including ulcers and salivary gland stones~\cite{liao2023ufct,liao2025imast}.
The acquisition of the oral and skin-hand volumes was approved by the University of Dundee School of Science and Engineering Research Ethics Committee (UOD-SSREC-PGR-2022-001 and UOD-SSREC-PGR-2022-003) and followed the Declaration of Helsinki, and all volunteers gave prior informed consent~\cite{liao2025imast,liao2023irunet}.
\begin{table}[!t]
  \caption{Datasets and their partitions.}
  \label{tab:data}
  \centering
  \begingroup
  \fontsize{7.6}{9.0}\selectfont
  \begin{threeparttable}
  \setlength{\tabcolsep}{2.2pt}
  \begin{tabular*}{\columnwidth}{@{\extracolsep{\fill}}lllccc@{}}
    \toprule
    Dataset & System & Probe & Training & Validation & Test \\
    \midrule
    Oral & SSOCT-1 & Intraoral & 111\,/\,6 & 15\,/\,8 & 26\,/\,3 \\
     & SSOCT-1 & LSM03 & 1\,/\,0 & 0\,/\,0 & 0\,/\,5 \\
     & SSOCT-2 & LSM03 & 44\,/\,0 & 11\,/\,0 & 15\,/\,1 \\
    \multicolumn{3}{@{}l}{\quad Total} & 162 & 34 & 50 \\
    \multicolumn{3}{@{}l}{\quad B-scans} & 74,400 & 15,800 & 24,200 \\
    \multicolumn{3}{@{}l}{\quad With half references} & 93 & 28 & 34 \\
    \addlinespace[2pt]
    Skin-hand & SSOCT-2 & LSM03 & -- & -- & 10 \\
    \addlinespace[2pt]
    OCTA-500 3M & -- & -- & 140 & 20 & 40 \\
    OCTA-500 6M & -- & -- & 200 & 30 & 70 \\
    \bottomrule
  \end{tabular*}
  \begin{tablenotes}[flushleft]
  \item[] Oral rows by system and probe give healthy\,/\,disease volumes. The skin-hand dataset has no training or validation set. OCTA-500 rows give subjects.
  \end{tablenotes}
  \end{threeparttable}
  \endgroup
\end{table}%

Two lab-built, hand-held swept-source (SS)-OCT systems, combined with two sample probes (the LSM03 objective and an intraoral probe~\cite{zhang2023probe}), were used to acquire the oral and skin-hand datasets.
SSOCT-1 has an A-scan rate of 200~kHz, a center wavelength of 1310~nm, a bandwidth of 100~nm and an in-air axial resolution and sampling of approximately 7.4 and 8.74~µm~\cite{liao2023irunet,liao2025imast}.
SSOCT-2 has an A-scan rate of 400~kHz, a center wavelength of 1300~nm and the same bandwidth~\cite{zhang2024rupp}.

On SSOCT-1, the LSM03 protocol uses \(600\times600\times960\) voxels (\(X,Y,Z\)), a \(5.16\times5.16\)~mm field of view and approximately 8.6~µm lateral spacing, and the intraoral protocol uses \(400\times400\times960\) voxels, a \(5.2\times5.2\)~mm field and approximately 13~µm lateral spacing~\cite{liao2025imast}.
Volumes used 3--12 repeats depending on session and probe.
SSOCT-2 was always combined with the LSM03 objective and acquired \(600\times600\times384\) voxels. The LSM03 objective images the labial mucosa, and the intraoral probe images the labial mucosa, floor of the mouth, tongue, buccal mucosa and gingiva~\cite{liao2025imast}.

The OCTA references \(T\) are computed from the complex OCT signals by eigen-decomposition~\cite{liao2024vet} in its windowed form~\cite{zhang2023wed} (axial window of length 10, step five).
Full and half references use separate decompositions, which discard rank \(r(n)=1,2,3\) for \(n<3\), \(3\leq n\leq5\) and \(n>5\) processed repeats.
\(X\) and \(T\) are in dB (\(20\log_{10}\) of amplitude).

All test volumes are isolated from training to prevent data leakage.
The training, validation and test sets are separated by participant, and all data were anonymized.
Two groups of test volumes evaluate generalization to conditions unseen in training.
The first group comprises the six LSM03 lesion volumes (ulcers and salivary gland stones) from three participants, and no LSM03 disease scan was used in training.
The second group is the skin-hand dataset, which comprises 10 volumes from five participants, acquired with SSOCT-2 and the LSM03 objective, with 12 repeats as in~\cite{liao2023irunet}.
All models keep their oral weights and inference settings.

For a fair comparison with published methods, ARCO was also trained and tested on OCTA-500 under the XOCT data separation protocol~\cite{xoct2025}, which fixes the training, validation and test subjects of the 3M (\(3\times3\)~mm) and 6M (\(6\times6\)~mm) subsets.%
\subsection{Implementation Details}\label{sec:implementation}

\(F\) is a U-Net regression network trained on the 162 training volumes with an MSE loss on the central B-scan of each seven-B-scan input, for 235,800 iterations with a batch size of 4.
It is frozen and used through its exponential moving average (EMA) weights.
The denoiser trains from random initialization for 707,400 iterations with AdamW (constant learning rate \(10^{-4}\), weight decay \(10^{-5}\)), a batch size of 4 and an EMA decay of 0.999.
The envelope loss is used on the 93 training volumes with half references, and the 69 three-repeat volumes use the bridge loss alone.
\(E_\phi\) is a U-Net without time conditioning, trained with \(L_E\) and \(L_{\mathrm{rank}}\) for 235,800 iterations with the same settings.
Training traverses all 943,200 fixed \(192\times192\) patches of the training volumes without replacement.

All networks are implemented in PyTorch 2.10 and operate on volumes at native resolution without flattening or resizing, with intensities normalized by statistics of the training data. The networks operate on \(192\times192\) patches, which overlap by 96 pixels at inference and are averaged to form the full B-scan.
The number of sampling steps \(N\) and the stride were chosen on the validation volumes, and oral predictions use \(N=8\) at a stride of 5 B-scans with \(\gamma=1\). ARCO has 104.6~M parameters, and a
\(400\times400\times960\) volume takes 136.6~s (about 2.3~min) on one RTX~5090.

On OCTA-500, data preprocessing follows XOCT~\cite{xoct2025}.
The denoiser, \(F\) and \(E_\phi\) are trained from scratch on the training subjects of each subset with the same optimizer settings and evaluated with final EMA weights. The envelope loss is omitted because OCTA-500 has no half references, and \(E_\phi\) takes the released inner limiting membrane (ILM) boundary as the tissue surface.
Sampling uses \(N=4\) at a stride of one B-scan with \(\gamma=1\) and averages four sampling runs, a setting chosen on the validation subjects.%
\subsection{Comparison Studies}\label{sec:arms}

Three variants of ARCO are trained for the same 707,400 iterations as the denoiser (Table~\ref{tab:test50}).
ARCO without the envelope loss (ARCO w/o \(L_{\mathrm{env}}\)) is trained with the bridge loss alone, and a second variant also replaces the HDiT denoiser with a convolutional U-Net denoiser.
The third variant, the deterministic denoiser, has the architecture and condition \(C\) of the denoiser and is trained at \(t=1\) alone, where the state equals \(A\), so that it reconstructs in a single sampling step.
The effect of the gate is measured by comparing ARCO at \(\gamma=1\) with the ungated bridge (\(\gamma=\infty\)).

On the oral dataset, ARCO is compared with seven published methods.
Four are OCT-to-OCTA translators, based on regression (VET~\cite{liao2024vet}), adversarial training (XOCT~\cite{xoct2025} and TransPro~\cite{transpro2024}) and flow matching (FlowOCT~\cite{flowoct2026}).
Three are diffusion models, developed for medical image translation (SelfRDB~\cite{arslan2025selfrdb}) and for natural images (ResShift~\cite{yue2025resshift} and a Palette-style model~\cite{saharia2022palette}).
All are trained on the 162 training volumes with their released code and model configurations, adapted to the volumetric data.
The baselines use each training patch one to four times, and the denoiser of ARCO three times.
Every baseline runs at the inference settings recommended by its published version, except that the Palette-style model uses 100 sampling steps in place of its released 1,000.
All models are evaluated with their final weights (EMA where kept), without early stopping or validation-based selection.

On OCTA-500, the comparison uses XOCT with its released weights and the results reported in~\cite{xoct2025} for XOCT and TransPro.
VET, ResShift and SelfRDB are retrained on the training subjects of each subset with the training settings of their oral runs and evaluated with their final weights.
FlowOCT is trained with its published schedule and its released code, which selects the checkpoint of highest validation peak signal-to-noise ratio (PSNR).%
\begin{table*}[!t]
  \caption{Oral test set (50 volumes, 41 healthy and nine disease), scored once with the frozen pipeline.}
  \label{tab:test50}
  \centering
  \begingroup
  \fontsize{7.4}{8.8}\selectfont
  \begin{threeparttable}
  \setlength{\tabcolsep}{1.8pt}
  \renewcommand{\arraystretch}{1.0}
  \begin{tabular*}{\linewidth}{@{\extracolsep{\fill}}lcccccccccc@{}}
    \toprule
     & \multicolumn{4}{c}{B-scans} & \multicolumn{3}{c}{En-face projection} & & \multicolumn{2}{c}{ARCO $-$ model, B-scans (95\% CI)} \\
    \cmidrule(lr){2-5}\cmidrule(lr){6-8}\cmidrule(l){10-11}
    Model & H (\%) $\downarrow$ & C (\%) $\uparrow$ & U (\%) $\downarrow$ & MAE (dB) $\downarrow$ & H$_0$ (\%) $\downarrow$ & C$_0$ (\%) $\uparrow$ & C$_1$ (\%) $\uparrow$ & Time (min) $\downarrow$ & $\Delta$H (\%) & $\Delta$C (\%) \\
    \midrule
    ARCO & \textbf{6.4} & 84.9 & \textbf{0.91} & 3.67 & \textbf{23.0} & \textbf{75.3} & \textbf{86.0} & 2.3 & -- & -- \\
    \addlinespace[2pt]
    \multicolumn{11}{@{}l}{\emph{Control and ablation}} \\
    ARCO, $\gamma=0$ (anchor \(F\)) & 4.9 & 66.5 & 0.82 & 3.26 & 20.4 & 75.4 & 83.9 & 0.9 & 1.5 (1.1, 1.8) & 18.4 (13.7, 22.7) \\
    ARCO, $\gamma=2$ & 8.8 & 90.7 & 1.43 & 3.87 & 24.2 & 75.3 & 86.6 & -- & $-$2.4 ($-$3.6, $-$1.3) & $-$5.8 ($-$8.5, $-$3.6) \\
    ARCO, $\gamma=\infty$ (ungated) & 8.8 & 90.7 & 1.48 & 3.87 & 24.2 & 75.3 & 86.6 & -- & $-$2.4 ($-$3.7, $-$1.3) & $-$5.8 ($-$8.5, $-$3.6) \\
    Deterministic denoiser (\(t=1\)) & 4.4 & 66.9 & 0.74 & 3.22 & 18.1 & 76.0 & 83.7 & -- & 2.0 (1.7, 2.3) & 18.0 (13.2, 22.5) \\
    ARCO w/o \(L_{\mathrm{env}}\) & 6.4 & 84.9 & 0.91 & 3.67 & 22.9 & 75.2 & 86.0 & -- & 0.0 ($-$0.1, 0.0) & 0.0 ($-$0.2, 0.1) \\
    ARCO w/o \(L_{\mathrm{env}}\), U-Net denoiser & 6.5 & 83.6 & 0.93 & 3.64 & 22.9 & 75.2 & 85.5 & -- & $-$0.1 ($-$0.2, 0.0) & 1.3 (1.0, 1.7) \\
    \addlinespace[2pt]
    \multicolumn{11}{@{}l}{\emph{Baselines}} \\
    VET & 6.8 & 61.7 & 1.37 & \textbf{3.66} & 25.5 & 69.6 & 79.0 & 0.9 & $-$0.4 ($-$1.6, 0.6) & 23.2 (19.0, 27.3) \\
    XOCT & 12.3 & \textbf{90.6} & 2.50 & 4.48 & 28.8 & 71.4 & 84.0 & 10.8 & $-$5.9 ($-$7.8, $-$4.2) & $-$5.7 ($-$8.7, $-$3.1) \\
    TransPro & 20.1 & 85.7 & 5.84 & 5.55 & 49.3 & 49.6 & 79.4 & \textbf{0.2} & $-$13.7 ($-$17.6, $-$10.3) & $-$0.8 ($-$4.5, 2.6) \\
    ResShift & 18.7 & 86.3 & 5.66 & 6.10 & 43.0 & 54.7 & 82.3 & 1.6 & $-$12.3 ($-$15.8, $-$9.3) & $-$1.4 ($-$4.3, 1.5) \\
    SelfRDB & 11.1 & 62.9 & 2.63 & 4.05 & 45.3 & 58.8 & 82.6 & 16.0 & $-$4.7 ($-$7.0, $-$2.8) & 22.0 (18.3, 25.3) \\
    Palette-style & 24.1 & 80.3 & 8.32 & 6.10 & 51.6 & 53.9 & 83.5 & 45.4 & $-$17.7 ($-$22.3, $-$13.6) & 4.6 (0.7, 7.9) \\
    FlowOCT & 12.6 & 77.8 & 2.98 & 4.22 & 42.1 & 64.0 & 78.9 & \textbf{0.2} & $-$6.2 ($-$8.8, $-$4.0) & 7.1 (2.8, 11.1) \\
    \bottomrule
  \end{tabular*}
  \begin{tablenotes}[flushleft]
  \item[] H, hallucination; C, completeness; U, unsupported fraction, on the 34 test volumes with half references. H$_0$, C$_0$ and C$_1$ are the en-face metrics. $\Delta$ is ARCO minus the model, with its 95\% CI. Time is to reconstruct one \(400\times400\times960\) volume on one RTX 5090 (dashes, not timed). Bold marks the best value among ARCO and the baselines.
  \end{tablenotes}
  \end{threeparttable}
  \endgroup
\end{table*}%
\subsection{Evaluation}\label{sec:metrics}

The acquired OCTA reference was used as the reconstruction target for evaluation.
Hallucination measures the predicted skeleton that does not exist in this reference, and completeness measures the reference skeleton that the prediction recovers.
Both are scored in the 192 pixels below the tissue surface, where the reference and the prediction are each segmented by an Otsu threshold and skeletonized, so that no vessel annotation is required.
With skeletons \(S_P,S_T\), masks \(M_P,M_T\) and a dilation \(\mathcal D_\tau\) by a tolerance of \(\tau\) pixels, hallucination (\(H_\tau\)) and completeness (\(C_\tau\)) are
  \begin{align}
    H_\tau&=\frac{|S_P\setminus\mathcal D_\tau(M_T)|}{|S_P|}, \label{eq:metric-hallucination}\\
    C_\tau&=1-\frac{|S_T\setminus\mathcal D_\tau(M_P)|}{|S_T|}. \label{eq:metric-completeness}
  \end{align}
For volumes with half references, with masks \(M_a,M_b\), the unsupported fraction (\(U_\tau\)) is the part of the predicted skeleton that neither half supports,
  \begin{equation}
    \label{eq:metric-unsupported}
    U_\tau=\frac{|S_P\setminus\mathcal D_\tau(M_a\cup M_b)|}{|S_P|}.
  \end{equation}
On every B-scan, hallucination (H), completeness (C) and the unsupported fraction (U) are computed with \(\tau=1\).

As secondary metrics, the same definitions are applied to the en-face mean projection of the 96 pixels below the tissue surface, with hallucination at zero tolerance (H\(_0\)) and completeness at zero and one pixel (C\(_0\), C\(_1\)).
Vessel density is the fraction of this projection covered by the mask (\(M_P\) or \(M_T\)).
The mean projections of the three 32-pixel slabs of this depth range are scored in the same way, to resolve the en-face metrics in depth.
Mean absolute error (MAE) is utilized as the pixel-wise error metric.
On OCTA-500, the released XOCT evaluation code is used, with full-projection (ILM to Bruch's membrane, BM) structural similarity index (SSIM) and PSNR as the primary metrics, and H and C are computed on the ILM--BM slab relative to the released OCTA reference.

Differences between models are reported with percentile 95\% confidence intervals (CIs) from a paired bootstrap with 10,000 resamples of identity blocks, each of which holds the volumes that share an identity label.%
\section{Results}\label{sec:results}

\subsection{Comparison on the Oral Test Set}\label{sec:test-results}

On the B-scans of the oral test set (Table~\ref{tab:test50}), ARCO reached 84.9\% completeness at 6.4\% hallucination and 3.67~dB MAE.
Across four sampling seeds, the mean completeness and hallucination varied by approximately 0.01\%.
Compared with the regression baseline VET, ARCO had 23.2\% higher completeness (95\% CI, 19.0--27.3) at similar hallucination (6.4\% vs.\ 6.8\%).
TransPro and ResShift had similar completeness to ARCO (differences within 1.4\%, 95\% CIs including zero) but with 13.7\% and 12.3\% more hallucination.
XOCT recovered 5.7\% more completeness but had 5.9\% more hallucination.
With the control value raised to \(\gamma=2\), ARCO achieved the 90.6\% completeness of XOCT ($\Delta$C 0.1\%; 95\% CI, \(-\)0.9 to 1.2) with 3.5\% less hallucination (95\% CI, 2.5--4.7) and 0.61~dB lower MAE (95\% CI, 0.47--0.76).
FlowOCT, SelfRDB and the Palette-style model performed worse than ARCO on all four B-scan metrics.
ARCO also had a lower unsupported fraction than each of the seven baselines (0.91\% vs.\ 1.37\%--8.32\%, every 95\% CI excluding zero) on the 34 test volumes with half references.
No baseline had both higher completeness and lower hallucination than ARCO on any test volume.

\begin{figure*}[!tp]
  \centering
  \includegraphics[width=\textwidth]{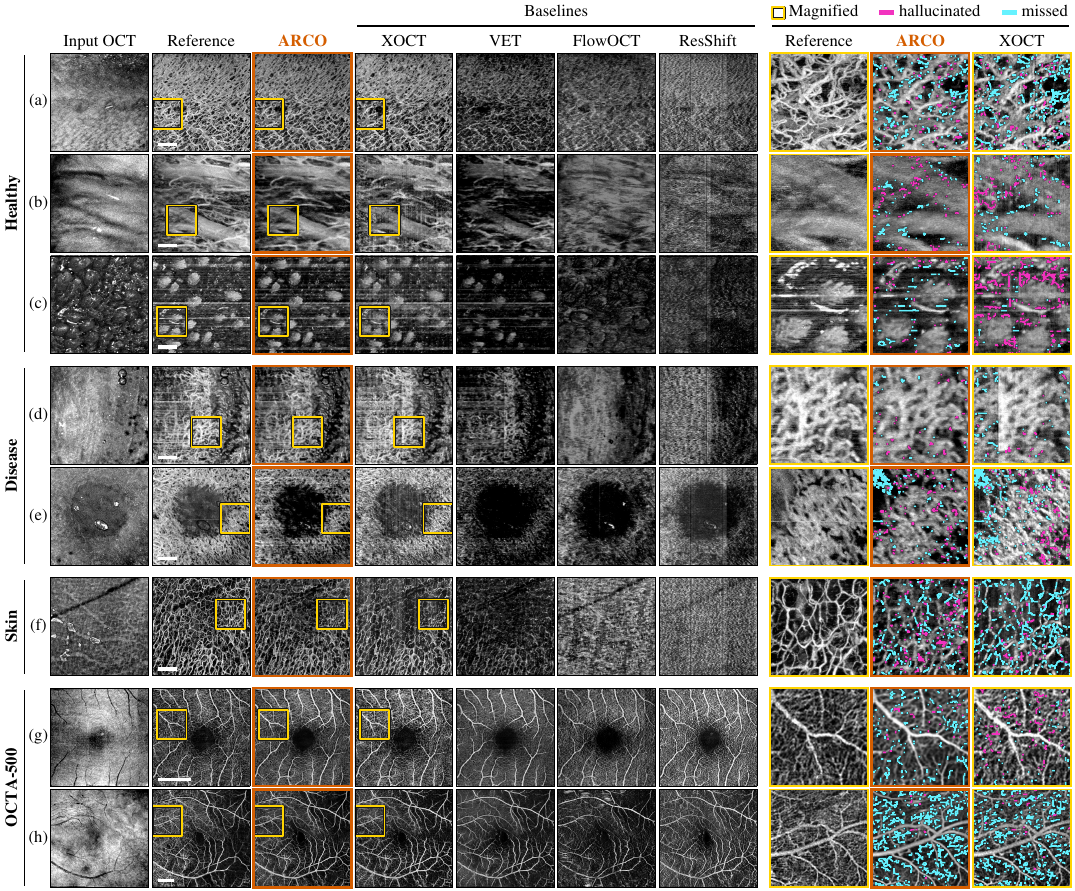}%
  \caption{%
    En-face examples of the oral, skin-hand and OCTA-500 test sets. (a)--(c) Healthy lip (LSM03 objective), floor of the mouth and tongue (intraoral probe). (d), (e) Disease, a mucosal lesion (intraoral probe) and an ulcer (LSM03 objective). (f) Skin-hand, zero-shot. (g), (h) Retinal OCTA-500 3M (normal) and 6M (diabetic retinopathy). Columns show the \singlescan{} input, the reference, ARCO (\(\gamma=1\)) and four baselines, with one display window per row. The right block magnifies the yellow window and marks hallucinated skeleton in magenta and missed skeleton in cyan. White bars, 1~mm.%
  }
  \label{fig:examples}
\end{figure*}

In the secondary en-face metrics, ARCO had higher completeness than every baseline, in C\(_0\) (75.3\% vs.\ 49.6\%--71.4\%) and C\(_1\) (86.0\% vs.\ 78.9\%--84.0\%), while keeping the lowest hallucination H\(_0\) (23.0\% vs.\ 25.5\%--51.6\%).
Every 95\% CI excluded zero except for C\(_1\) compared to the Palette-style model.
Compared with XOCT (the baseline with the highest B-scan completeness), ARCO had 3.9\% higher completeness C\(_0\) (95\% CI, 2.5--5.1) with 5.8\% lower hallucination H\(_0\) (95\% CI, 4.0--8.0).
The vessel density of the ARCO projection agreed with the reference (\(r=0.91\) vs.\ 0.06--0.75 for the baselines; absolute error, 2.7\%), with a lower error than all baselines.
Fig.~\ref{fig:examples} shows en-face examples of the test sets, where the magnified panels mark hallucinated skeleton in magenta and missed skeleton in cyan for ARCO and XOCT.

\subsection{Disease and Skin-Hand Data}\label{sec:skin-results}

On the nine disease volumes (two examples in Fig.~\ref{fig:examples}d, e), ARCO kept its completeness (84.5\% vs.\ 85.0\% on the healthy volumes) with lower hallucination than all seven baselines (11.5\% vs.\ 16.2\%--36.1\%, Table~\ref{tab:strata}).
Its hallucination was lower than for every baseline on eight of the nine volumes.
On the six LSM03 scans with oral lesions, a condition unseen in training, ARCO had 10.9\% hallucination, compared with 17.5\%--36.2\% for the baselines.

\begin{table}[!b]
  \caption{B-scan hallucination (H) and completeness (C) by health group and on the skin-hand test set (\%).}
  \label{tab:strata}
  \centering
  \begingroup
  \fontsize{7.6}{9.0}\selectfont
  \begin{threeparttable}
  \setlength{\tabcolsep}{2.2pt}
  \begin{tabular*}{\columnwidth}{@{\extracolsep{\fill}}lcccccc@{}}
    \toprule
     & \multicolumn{2}{c}{Healthy (41)} & \multicolumn{2}{c}{Disease (9)} & \multicolumn{2}{c}{Skin-hand (10)} \\
    \cmidrule(lr){2-3}\cmidrule(lr){4-5}\cmidrule(l){6-7}
    Model & H $\downarrow$ & C $\uparrow$ & H $\downarrow$ & C $\uparrow$ & H $\downarrow$ & C $\uparrow$ \\
    \midrule
    ARCO & 5.3 & 85.0 & \textbf{11.5} & 84.5 & \textbf{23.3} & 71.3 \\
    ARCO, $\gamma=0$ (anchor \(F\)) & 3.8 & 65.6 & 9.7 & 70.4 & 28.3 & 71.6 \\
    ARCO, $\gamma=\infty$ (ungated) & 6.7 & 90.3 & 18.3 & 92.5 & 36.2 & 94.7 \\
    \addlinespace[2pt]
    VET & \textbf{4.7} & 58.8 & 16.2 & 74.8 & 40.1 & 62.1 \\
    XOCT & 10.1 & \textbf{90.1} & 22.4 & \textbf{92.8} & 42.1 & \textbf{92.7} \\
    TransPro & 18.6 & 84.7 & 27.1 & 90.0 & 47.2 & 88.3 \\
    ResShift & 15.5 & 85.6 & 33.6 & 89.4 & 45.2 & 74.4 \\
    SelfRDB & 8.6 & 61.1 & 22.2 & 71.0 & 38.3 & 53.1 \\
    Palette-style & 21.5 & 79.5 & 36.1 & 84.4 & 54.8 & 78.0 \\
    FlowOCT & 11.4 & 76.2 & 18.0 & 84.8 & 41.8 & 72.0 \\
    \bottomrule
  \end{tabular*}
  \begin{tablenotes}[flushleft]
  \item[] Scoring as in Table~\ref{tab:test50}. Six of the nine disease volumes are LSM03 lesion scans, and no LSM03 disease scan entered training. Skin-hand, 10 volumes from five independent participants, every model at its oral weights and settings. Bold marks the best value of each column among ARCO and the seven baselines.
  \end{tablenotes}
  \end{threeparttable}
  \endgroup
\end{table}%

In zero-shot inference on the skin-hand volumes (Fig.~\ref{fig:examples}f), ARCO had lower hallucination than its anchor and all seven baselines on each volume.
Its mean hallucination was 23.3\%, compared with 38.3\%--54.8\% for the baselines.
Relative to its anchor, ARCO had similar completeness ($\Delta$ \(-\)0.3\%; 95\% CI, \(-\)1.2 to 0.7) with 5.0\% less hallucination (95\% CI, 3.8--6.1).
With the gate removed (\(\gamma=\infty\)), ARCO had higher completeness (94.7\% vs.\ 92.7\%) and lower hallucination (36.2\% vs.\ 42.1\%) than XOCT on all 10 volumes.
XOCT and TransPro recovered 21.4\% and 17.0\% more completeness than ARCO at \(\gamma=1\), but with 1.8 and 2.0 times its hallucination.
These results indicate that the hallucination control of ARCO is transferable to unseen lesions and to another anatomical site.%
\subsection{Retinal OCTA-500}\label{sec:octa500-results}

\begin{table*}[!tp]
  \caption{Retinal OCTA-500 comparison under the XOCT protocol, with separate models for the 3M and 6M subsets.}
  \label{tab:octa500}
  \centering
  \begingroup
  \fontsize{7.2}{8.6}\selectfont
  \begin{threeparttable}
  \setlength{\tabcolsep}{2.0pt}
  \renewcommand{\arraystretch}{1.0}
  \begin{tabular*}{\linewidth}{@{\extracolsep{\fill}}lcccccccccccc@{}}
    \toprule
     & \multicolumn{4}{c}{Full projection (ILM--BM)} & \multicolumn{4}{c}{Mean projection} & \multicolumn{3}{c}{3D volume} & C\,/\,H (\%, ILM--BM) \\
    \cmidrule(lr){2-5}\cmidrule(lr){6-9}\cmidrule(lr){10-12}\cmidrule(l){13-13}
    Method & MAE$\downarrow$ & PSNR$\uparrow$ & SSIM$\uparrow$ & Perp$\downarrow$ & MAE$\downarrow$ & PSNR$\uparrow$ & SSIM$\uparrow$ & Perp$\downarrow$ & MAE$\downarrow$ & PSNR$\uparrow$ & SSIM$\uparrow$ & $\uparrow$\,/\,$\downarrow$ \\
    \midrule
    \multicolumn{13}{@{}l}{\emph{OCTA-3M (40 test subjects)}} \\
    TransPro~\cite{transpro2024}$^{a}$ & 19.54 & 20.14 & 0.580 & 0.608 & 20.93 & 19.83 & 0.573 & 0.582 & 3.30$^{c}$ & 28.46 & 0.866 & -- \\
    XOCT~\cite{xoct2025}$^{a}$ & 19.22 & 20.21 & 0.608 & 0.573 & 20.54 & 19.90 & 0.596 & \textbf{0.558} & 2.78 & 30.08 & 0.893 & -- \\
    XOCT, released weights$^{b}$ & 19.12 & 20.27 & 0.610 & \textbf{0.572} & 20.55 & 19.90 & 0.596 & \textbf{0.558} & 2.78 & 30.08 & 0.893 & 79.9\,/\,19.4 \\
    FlowOCT~\cite{flowoct2026}, retrained$^{d}$ & 22.86 & 18.97 & 0.455 & 0.683 & 24.75 & 18.56 & 0.444 & 0.653 & 2.97 & 30.01 & 0.872 & \textbf{82.7}\,/\,25.1 \\
    VET~\cite{liao2024vet}, retrained$^{e}$ & 17.84 & 21.03 & 0.590 & 0.615 & 20.21 & 20.29 & 0.569 & 0.600 & 2.46 & 31.31 & 0.906 & 80.3\,/\,15.6 \\
    ResShift~\cite{yue2025resshift}, retrained$^{e}$ & 24.24 & 18.41 & 0.482 & 0.707 & 30.04 & 17.14 & 0.456 & 0.676 & 3.68 & 29.15 & 0.788 & 74.1\,/\,20.8 \\
    SelfRDB~\cite{arslan2025selfrdb}, retrained$^{e}$ & 20.20 & 19.85 & 0.524 & 0.664 & 21.99 & 19.46 & 0.513 & 0.634 & 2.58 & 30.71 & 0.894 & 72.0\,/\,16.5 \\
    \textbf{ARCO} & \textbf{17.22} & \textbf{21.29} & \textbf{0.626} & 0.609 & \textbf{19.16} & \textbf{20.68} & \textbf{0.606} & 0.594 & \textbf{2.42} & \textbf{31.49} & \textbf{0.908} & 82.3\,/\,\textbf{15.4} \\
    \addlinespace[2pt]
    \multicolumn{13}{@{}l}{\emph{OCTA-6M (70 test subjects)}} \\
    TransPro~\cite{transpro2024}$^{a}$ & 17.40 & 21.07 & 0.524 & 0.591 & 17.42 & 21.24 & 0.539 & 0.570 & 2.72$^{c}$ & 29.92 & 0.887 & -- \\
    XOCT~\cite{xoct2025}$^{a}$ & 16.65 & 21.46 & 0.568 & \textbf{0.541} & 16.21 & 21.85 & 0.574 & \textbf{0.537} & 2.34 & 31.26 & 0.905 & -- \\
    XOCT, released weights$^{b}$ & 16.59 & 21.51 & 0.570 & \textbf{0.541} & 16.21 & 21.86 & 0.574 & \textbf{0.537} & 2.34 & 31.26 & 0.902 & 76.9\,/\,18.3 \\
    FlowOCT~\cite{flowoct2026}, retrained$^{d}$ & 20.00 & 19.80 & 0.422 & 0.665 & 20.07 & 19.96 & 0.407 & 0.650 & 2.65 & 30.22 & 0.887 & 74.7\,/\,24.1 \\
    VET~\cite{liao2024vet}, retrained$^{e}$ & 15.03 & 22.32 & 0.553 & 0.613 & 15.30 & 22.33 & 0.553 & 0.591 & 2.09 & 32.29 & 0.913 & \textbf{79.5}\,/\,16.3 \\
    ResShift~\cite{yue2025resshift}, retrained$^{e}$ & 18.88 & 20.46 & 0.466 & 0.673 & 18.88 & 20.64 & 0.460 & 0.644 & 2.66 & 30.56 & 0.881 & 77.5\,/\,21.2 \\
    SelfRDB~\cite{arslan2025selfrdb}, retrained$^{e}$ & 20.05 & 19.95 & 0.491 & 0.640 & 20.51 & 19.98 & 0.505 & 0.613 & 2.26 & 31.40 & 0.898 & 67.1\,/\,\textbf{13.6} \\
    \textbf{ARCO} & \textbf{14.79} & \textbf{22.46} & \textbf{0.582} & 0.607 & \textbf{15.22} & \textbf{22.39} & \textbf{0.585} & 0.586 & \textbf{2.06} & \textbf{32.38} & \textbf{0.915} & \textbf{79.5}\,/\,14.3 \\
    \bottomrule
  \end{tabular*}
  \begin{tablenotes}[flushleft]
  \item[] Metrics follow the released XOCT evaluation method, with MAE on 0--255 intensities, PSNR in dB and Perp, the perceptual distance. The full projection averages the slab between the inner limiting membrane (ILM) and Bruch's membrane (BM), and the mean projection averages the full depth. C\,/\,H is the B-scan completeness and hallucination on the ILM--BM slab. Bold marks the best value in each subset. Differences in the text use unrounded values. $^{a}$Reported in~\cite{xoct2025}. $^{b}$Evaluated with the released code. $^{c}$Reported in~\cite{xoct2025} as inflated by a uint8 overflow. $^{d}$Retrained with its published schedule. $^{e}$Retrained with the settings of its oral run.
  \end{tablenotes}
  \end{threeparttable}
  \endgroup
\end{table*}%

On OCTA-500 (Table~\ref{tab:octa500} and Fig.~\ref{fig:examples}g, h), ARCO outperformed XOCT and TransPro in full-projection SSIM and PSNR.
On 3M, ARCO reached an SSIM of 0.626 and a PSNR of 21.29~dB, compared with 0.610 and 20.27~dB for XOCT, an improvement of 0.016 (95\% CI, 0.008--0.023) and 1.01~dB (95\% CI, 0.77--1.26).
On 6M, it reached 0.582 and 22.46~dB, compared with 0.570 and 21.51~dB, with differences of 0.013 (95\% CI, 0.010--0.016) and 0.95~dB (95\% CI, 0.88--1.02).
In the mean projection and in the 3D volume, ARCO also had the lowest MAE and the highest PSNR and SSIM of all models.
Its perceptual distance was larger than for XOCT and TransPro.

The released XOCT weights reproduced the published full-projection XOCT results to within 0.002 SSIM and 0.06~dB PSNR.
The retrained FlowOCT had a 3D SSIM of 0.872 on 3M and 0.887 on 6M, compared with 0.873 and 0.871 in~\cite{flowoct2026}.

ARCO also outperformed the retrained FlowOCT, VET, ResShift and SelfRDB in full-projection SSIM and PSNR on both subsets.
Among these models, the smallest improvement was over VET on 6M, with 0.029 in SSIM (95\% CI, 0.027--0.032) and 0.13~dB in PSNR (95\% CI, 0.09--0.18).
On the ILM--BM slab of 3M, ARCO had 2.4\% higher B-scan completeness (95\% CI, 1.5--3.2) and 4.0\% lower hallucination (95\% CI, 3.1--4.8) than XOCT.
On 6M, it had 2.5\% higher completeness (95\% CI, 1.4--3.6) and 3.9\% lower hallucination (95\% CI, 3.4--4.5).
These results indicate that the design of ARCO also applies to retinal OCTA.%
\subsection{Control and Ablation}\label{sec:control-results}

\begin{figure}[!t]
  \centering
  \includegraphics[width=\columnwidth]{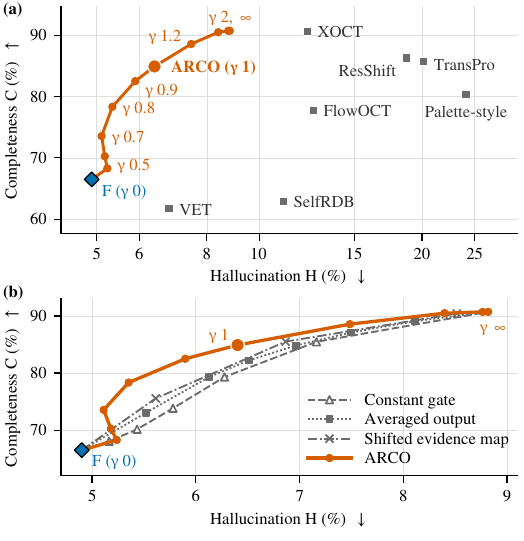}%
  \caption{%
    B-scan completeness and hallucination on the oral test set. (a) The orange curve shows ARCO performance at \(\gamma=0\) (the anchor \(F\), blue diamond), 0.25, 0.5, 0.6, 0.7, 0.8, 0.9, 1, 1.2, 1.5, 2 and \(\infty\) (ungated). The points at \(\gamma=0\) and 0.25 overlap, as do those at \(\gamma=2\) and \(\infty\). Gray squares are the baseline results from Table~\ref{tab:test50}. The hallucination axis is logarithmic. (b) Mean completeness and hallucination of ARCO and the three simpler gates from the anchor to the ungated bridge, on linear axes.%
  }
  \label{fig:gamma}
\end{figure}

Increasing \(\gamma\) from 0 to \(\infty\) raised completeness from 66.5\% to 90.7\% and hallucination from 4.9\% to 8.8\% on the oral test set (Fig.~\ref{fig:gamma}a).
The completeness of ARCO increased monotonically with \(\gamma\), and each baseline had more hallucination than an ARCO setting of equal or higher completeness.

Compared with the regression anchor \(F\), ARCO had 18.4\% higher completeness (95\% CI, 13.7--22.7) with 1.5\% more hallucination (95\% CI, 1.1--1.8).
In the en-face metrics, ARCO had similar completeness to its anchor in C\(_0\) (75.3\% vs.\ 75.4\%) and higher completeness in C\(_1\) (86.0\% vs.\ 83.9\%), with higher hallucination H\(_0\) (23.0\% vs.\ 20.4\%).
The increase in C\(_1\) was largest in the 32 pixels nearest the surface (9.9\%).

Removing the gate (\(\gamma=\infty\)) added 5.8\% completeness (95\% CI, 3.6--8.5) at 2.4\% more hallucination (95\% CI, 1.3--3.7).
At \(\gamma=1\), ARCO thus kept most of the completeness improvement of the ungated bridge over the anchor for less than half of the added hallucination.
The effect of the gate grew from the healthy to the disease and the skin-hand volumes (Table~\ref{tab:strata}).
Removing it raised hallucination from 5.3\% to 6.7\% on the healthy, from 11.5\% to 18.3\% on the disease and from 23.3\% to 36.2\% on the skin-hand volumes.

The evidence gate was further tested by replacing it with three simpler gates on the oral test set.
The first used one constant value for every pixel, and the second averaged the anchor and the ungated output with a fixed weight.
The third used the evidence map of ARCO shifted by half the field.
Each was compared with ARCO at the same B-scan completeness, by interpolation over its settings in each volume (Fig.~\ref{fig:gamma}b).
At that completeness, ARCO had less hallucination than each of them, by 0.5\% (95\% CI, 0.3--0.8), 0.4\% (95\% CI, 0.2--0.6) and 0.4\% (95\% CI, 0.3--0.6).
On the nine disease volumes the differences were 0.9\%--1.6\%, and ARCO had the lower hallucination on eight of the nine in each comparison.
The shifted gate applies the same evidence values at shifted locations, and its higher hallucination confirms that the location of the evidence contributes to hallucination control.

Three variants of ARCO (Table~\ref{tab:test50}) were tested for the contribution of the iterative sampling, the denoiser architecture and the envelope loss.
The deterministic denoiser, which reconstructs in one sampling step, had 18.0\% less B-scan completeness than ARCO (95\% CI, 13.2--22.5).
In the en-face metrics, it had lower H\(_0\) (18.1\% vs.\ 23.0\%) and higher C\(_0\) (76.0\% vs.\ 75.3\%) than ARCO, so the iterative correction did not improve en-face metrics at zero tolerance.
Compared with a U-Net denoiser trained in the same way, the denoiser of ARCO recovered 1.3\% more completeness (95\% CI, 1.0--1.8), with no measurable change in hallucination (difference \(-\)0.1\%; 95\% CI, \(-\)0.2 to 0.0).
Removing the envelope loss changed the B-scan metrics slightly.%
\section{Discussion}\label{sec:discussion}

This work presents ARCO, a regression-anchored controllable diffusion model for \singlescan{} OCT-to-OCTA translation.
For clinical use, OCTA reconstructed from a single scan should have high vessel completeness and low hallucination, because missed or hallucinated vessels could mislead the assessment of the microvasculature.
On the oral test set, no baseline achieved both higher completeness and lower hallucination than ARCO, and a single control value \(\gamma\) adjusted this trade-off from the anchor to the ungated bridge without retraining.
On the retinal OCTA-500 dataset, the retrained ARCO outperformed XOCT.
The hallucination control of ARCO transferred to oral lesions unseen in training and, in zero-shot inference, to skin-hand data.
This transferability is important for clinical use, since lesion vasculature can differ from the healthy vasculature in most of the training volumes.

Two ablation studies identify the source of the improvement over the anchor.
Replacing the iterative sampling with a single-step deterministic denoiser reduced B-scan completeness by 18.0\%, which confirms the contribution of the bridge and its iterative sampling.
However, this replacement did not degrade the en-face metrics H\(_0\) and C\(_0\), in agreement with the performance reported for regression models in other medical image translation tasks~\cite{rassmann2026yoda}.
In addition, replacing the evidence gate with any of the three simpler gates, including the shifted evidence map, increased hallucination at the same completeness.
The evidence gate is therefore effective through the spatial localization of its evidence.

Compared with ARCO, the regression baseline VET had similarly low hallucination but lower completeness, and XOCT, TransPro and ResShift achieved similar or higher completeness but at approximately two to three times the hallucination.
Similar to predict-and-refine models~\cite{whang2022stochastic,ohayon2025pmrf,chen2025madm}, ARCO starts from a regression output, and it further introduces a spatially adaptive gate that weights the refinement by the estimated pixel-wise error of this output.
Because \(\gamma\) is applied during sampling, a single trained model covers the range between the anchor and the ungated bridge (Fig.~\ref{fig:gamma}a), and the trade-off can be set for each application.

The transferability of the hallucination control is attributed to the evidence gate.
The evidence model predicts the error of the anchor from the \singlescan{} structural OCT, and the gate restricts the correction to the regions of large predicted error.
Outside these regions, the output stays close to the regression anchor, including on data unseen in training.
Without this restriction, the ungated bridge applies the correction globally, and its hallucination increased most on the skin-hand data (Table~\ref{tab:strata}).
With it, ARCO had lower hallucination than the regression baseline VET on the disease and skin-hand volumes, and its completeness was higher.
ARCO therefore recovers more vessels than this regression baseline without adding hallucination on data unseen in training.
On OCTA-500, the improvement over XOCT further indicates that ARCO is not specific to the oral data or to the metrics defined in this study.
ARCO could be extended to routine retinal imaging, where OCT is used more widely than OCTA and most clinics hold extensive OCT databases~\cite{lee2019flow}.
Such an extension would provide vascular information without an additional scan.

This study has limitations.
First, the acquired OCTA reference is not a ground truth of the vasculature, and hallucination and completeness are defined relative to this reference.
While the unsupported fraction computed from both half references mitigates this dependence, future work will consider utilizing a phantom suitable for micrometer-scale OCTA imaging as a gold standard.
Second, the number of disease volumes is limited in this study.
A collaboration with clinical oral medicine is being established, and future work will optimize ARCO for oral diseases, including oral cancer.%
\needspace{3\baselineskip}
\section{Conclusion}\label{sec:conclusion}

ARCO demonstrates that hallucination in OCT-to-OCTA translation can be controlled during sampling by gating the diffusion correction of a regression anchor with its predicted error.
This control balanced vessel completeness and hallucination on oral data, where ARCO had higher completeness than the regression baseline and lower hallucination than the baselines of similar completeness.
On unseen oral lesions and, in zero-shot inference, on skin-hand data, ARCO also had lower hallucination than every baseline.
Retrained on retinal OCTA-500, ARCO outperformed the state-of-the-art XOCT under the same experimental protocol.
These results indicate the potential of ARCO to provide vascular images from a single fast scan in oral and retinal imaging limited by slow repeated scanning.%

\bibliographystyle{IEEEtran}
\bibliography{references}
\end{document}